\documentclass[%
preprint,
superscriptaddress,
showpacs,preprintnumbers,
 amsmath,
 amssymb,
aps,
pre,
floatfix,
]{revtex4-1}
\usepackage{graphicx}
\usepackage[colorlinks,allcolors=blue]{hyperref}
\usepackage{amsmath,amssymb,amsthm}

\usepackage{array, booktabs, makecell}

\begin{document}

\title{The role of fractional time-derivative operators on anomalous diffusion}

\author{Angel A. Tateishi}
\affiliation{Departamento de F\'{\i}sica, Universidade Tecnol\'ogica Federal do Paran\'a, Pato Branco, PR 85503-390, Brazil}
\author{Haroldo V. Ribeiro}
\affiliation{Departamento de F\'isica, Universidade Estadual de Maring\'a, Maring\'a, PR 87020-900, Brazil}
\author{Ervin K. Lenzi}\email{eklenzi@uepg.br}
\affiliation{Departamento de F\'isica, Universidade Estadual de Ponta Grossa, Ponta Grossa, PR 84030-900, Brazil}

\linespread{1.4}
\begin{abstract}
The generalized diffusion equations with fractional order derivatives have shown be quite efficient to describe the diffusion in complex systems, with the advantage of producing exact expressions for the underlying diffusive properties. Recently, researchers have proposed different fractional-time operators (namely: the Caputo-Fabrizio and Atangana-Baleanu) which, differently from the well-known Riemann-Liouville operator, are defined by non-singular memory kernels. Here we proposed to use these new operators to generalize the usual diffusion equation. By analyzing the corresponding fractional diffusion equations within the continuous time random walk framework, we obtained waiting time distributions characterized by exponential, stretched exponential, and power-law functions, as well as a crossover between two behaviors. For the mean square displacement, we found crossovers between usual and confined diffusion, and between usual and sub-diffusion. We obtained the exact expressions for the probability distributions, where non-Gaussian and stationary distributions emerged. This former feature is remarkable because the fractional diffusion equation is solved without external forces and subjected to the free diffusion boundary conditions. We have further shown that these new fractional diffusion equations are related to diffusive processes with stochastic resetting, and to fractional diffusion equations with derivatives of distributed order. Thus, our results show that these new operators are a simple and efficient way for incorporating different structural aspects into the system, opening new possibilities for modeling and investigating anomalous diffusive processes.

\end{abstract}
\maketitle

\section{Introduction}

The {\it random walk} concept is one of the broadest and versatile paradigms to deal with statistical fluctuations. The term ``random walk'' was coined in 1905 by Pearson~\cite{Pearson}, but the fundamental relationship between this concept and the usual diffusion equation was reported earlier in the seminal works of Rayleigh~\cite{Rayleigh1,Rayleigh2,Rayleigh3} in sound theory, Bachelier~\cite{Bachelier} in economics, Einstein~\cite{Einstein} and Smoluchowski~\cite{Smoluchowski} in the Brownian motion theory. Due to this intrinsic relation, the usual random walk is characterized by Gaussian, Markovian, and ergodic properties, which lead to a linear time dependence of the mean square displacement, $(\Delta x)^2 \sim t$. The versatility of this concept relies on the possibility of generalizations and extensions to describe systems with one or more characteristics of anomalous diffusion: non-Gaussian distributions; long-range memory effects (non-Markovian); non-ergodicity; divergent mean square displacement (L\'evy walks); and nonlinear mean square displacement, $(\Delta x)^2 \sim t^{\alpha}$ (sub-diffusion: $\alpha<1$, superdiffusion: $\alpha>1$, confined or saturated diffusion: $\alpha=0$).

In the context of generalizations, the first landmark is the work proposed in 1965 by Montroll and Weiss~\cite{Montroll}, in which they introduce the {\it continuous time random walk} concept (see Ref.~\cite{Kutner} for a general overview). This framework is characterized by a joint distribution of jump length and waiting time $\psi(x,t)$, where $\lambda(x)=\int_{0}^{\infty} dt \psi(x,t)$ is the jump length distribution and $\omega(t)=\int_{-\infty}^{\infty} dx \psi(x,t)$ is the waiting time distribution. Subsequently, a connection with a generalized master equation is proposed to discuss memory effects in the continuous time random walk~\cite{MontrollScher,Kenkre,Shlesinger}; the waiting time distribution $\omega(t)$ is strictly related to the memory kernel of the generalized master equation. Moreover, natural extensions of both random walk and continuous time random walk were proposed to study transport properties in systems with structural complexity, such as disordered, random, and fractal environments~\cite{Havlin2000}.

The second landmark, ``a modern era of the continuous time random walk'' according to Kutner and Masoliver~\cite{Kutner}, is the development of the intrinsic relationship between this formalism and the fractional diffusion equations. Among the seminal works, we have found a simple mention in the work of Klafter~\textit{et al.}~\cite{Klafter1987} on the possibility of having a fractional diffusion equation to describe anomalous transport. However, were Hilfer~\textit{et al.}~\cite{Hilfer1995} that, ninety years after Einstein's work~\cite{Einstein}, established a rigorous and precise connection between the continuous time random walk and the fractional master equation as well as with the fractional diffusion equation (a special case of the former). This result was later extended by Compte~\cite{Compte1996} in the long-time limit, where it is shown that any decoupled continuous time random walk having no characteristic scale of time or space (power-law memories) corresponds to a time- or space-fractional diffusion equation, respectively with Riemann-Liouville time derivative or Riesz space derivate. The subsequent success and development of the fractional approach are well documented in two review articles by Meztler and Klafter~\cite{15,Restaurant}, and in several articles by Barkai~\cite{Barkai1,Barkai2,Barkai3,Barkai4}.

Since memory effects underlying a continuous time random walk are implicitly considered by the differential operators, the versatility of the fractional formalism is mainly related to two remarkable features. First, this formalism handles very well the physical requirements of a system by dealing boundary conditions and external forces in a simple manner. Second, it takes advantage of traditional tools from mathematical physics and statistics for obtaining exact expressions to describe complex systems with anomalous behaviors. For instance, in the following fractional differential equation
\begin{eqnarray}
\frac{\partial}{\partial t}\rho(x,t) = \;_{0}{\cal{D}}_{t}^{1-\alpha}\left({\cal{L}}\{\rho(x,t)\}\right), 
\label{eq1}
\end{eqnarray}
where
\begin{eqnarray}
{\cal{L}}\{\rho(x,t)\}= D\frac{\partial^{2}}{\partial x^{2}}\rho(x,t)-\frac{\partial}{\partial x}[F(x,t)\rho(x,t)], \nonumber
\end{eqnarray}
the nonusual relaxation can be associated with a continuous time random walk where the waiting time distribution is a power-law. This process is also strictly related to the Riemann-Liouville fractional operator~\cite{BarkaiMetzlerKlafter}
\begin{eqnarray}
\;_{0}{\cal{D}}_{t}^{1-\alpha}\rho(x,t)=\frac{1}{\Gamma\left(\alpha\right)}\frac{d}{dt}\int_{0}^{t}\frac{\rho(x,t')}{(t-t')^{1-\alpha}}dt'\;,
\end{eqnarray} 
where $0<\alpha<1$ is the {\it fractional order exponent} (or the {\it anomalous exponent}), a quantity that can be interpreted as an index of memory in empirical systems~\cite{Du2013}. The fractional operator is also responsible for introducing a nonlinear time dependence in the mean square displacement of the system~\cite{15}. Thus, a large class of complex phenomena can be effectively described by extending the standard differential operator to a non-integer order~\cite{zahran,zahran2,zahran3,villa,burioni,maringa,iomin5,iomin6,maringa2,ChemPhys}; indeed, as pointed out by West~\cite{WestBook}, the fractional calculus provides a suitable framework to deal with complex systems. 

Recently, researchers have made and promoted remarkable progress toward improving experimental techniques for investigating diffusive processes, mainly illustrated by the developments in the single-particle tracking technique~\cite{Saxton1997, Wirtz2009,Gal2013,Hoze2017}. Such improvements yield novel insights into transport properties of biological systems~\cite{Franosch2013,Manzo2015,Shen2017} and nanomaterials~\cite{Zagato2014,Karger2012,Karger2016}, where the high-resolution of the experiments has found different diffusive behaviors depending on the time scale. In this context, an important question is whether other forms of fractional differential operators (replacing the Riemann-Liouville one) such as those recently-proposed with non-singular kernels~\cite{nonsingular1,singularJordan,GA,nonsingular2,AB,nonsingular3} are suitable to describe the aforementioned situations. To answer this question, we investigate an one-dimensional diffusive process described by the fractional diffusion equation
\begin{eqnarray}
\label{PAe1b}
\frac{\partial}{\partial t} \rho(x, t)=D {\cal{F}}_{t}^{\alpha}\left( \frac{\partial^2}{\partial x^2} \rho(x, t)\right)\,,
\end{eqnarray}
where $D$ is the generalized diffusion coefficient. This equation is also subjected to the free diffusion boundary conditions $\rho(\pm \infty,t)=0$ and to the initial condition $\rho(x,0)=\varphi(x)$. 

The fractional operator in Eq.~(\ref{PAe1b}) is defined as
\begin{eqnarray}
\label{KK}
{\cal{F}}_{t}^{\alpha}\rho(x,t)=\frac{\partial}{\partial t} \int_{0}^{t}\rho(x,t'){\cal{K}}(t-t')dt'\;,
\end{eqnarray}
in order to consider situations with singular and non-singular kernels in a unified way. It is worth noting that ${\cal{K}}(t)=\delta(t)$ recovers the usual diffusion equation. Here we consider three different forms for the kernel ${\cal{K}}(t)$. The first one is 
\begin{eqnarray}
\label{KK11}
{\cal{K}}(t)=  \frac{\left(t/\tau\right)^{\alpha-1}}{\Gamma\left(\alpha\right)}\,,
\end{eqnarray}
which corresponds to the well-known Riemann-Liouville fractional operator~\cite{Livro1} for $0<\alpha<1$. The second one is
\begin{eqnarray}\label{KK22e}
{\cal{K}}(t)&=& b\,e^{-\frac{\alpha t}{(1-\alpha)\tau}}\,,
\end{eqnarray}
which corresponds to the fractional operator of Caputo-Fabrizio~\cite{nonsingular1}. In the context of the diffusion equation, the use of this operator is related with a diffusion equation combined with first-order kinetics~\cite{Sano}. 

Finally, the third one is
\begin{eqnarray}\label{KK33}
{\cal{K}}(t)= b\,E_{\alpha}\left[-\frac{\alpha}{1-\alpha} \left(\frac{t}{\tau}\right)^{\alpha}\right]\,,
\end{eqnarray}
where $E_{\alpha}(\dots)$ is the Mittag-Leffler function~\cite{Livro1}. This kernel corresponds to the fractional operator of Atangana-Baleanu~\cite{AB}. Further possibilities for the kernel ${\cal{K}}(t)$ are discussed by G\'omez-Aguilar~\textit{et al.}~\cite{GA}. We observe that the Riemann-Liouville operator have a singularity at the origin ($t=0$), while the recently-proposed Caputo-Fabrizio and Atangana-Baleanu are non-singular operators~\cite{nonsingular1,singularJordan,GA,nonsingular2,AB,nonsingular3}. In the previous definitions, the parameter $b$ is a normalization constant, $\alpha$ is the fractional order exponent, and $\tau$ is a characteristic time that controls the shape of the kernels. 

Our main goal here is to verify how these different fractional operators modify the fractional diffusion equation~(\ref{eq1}) and what are the effect of these choices on the underlying diffusive properties of a system modeled by this equation. The rest of this manuscript is organized as follows. In Section~II, we investigate general solutions and processes related with Eq.~(\ref{PAe1b}) when considering different choices (singular and non-singular) for the kernel ${\cal{K}}(t)$. In Section III, we present a summary of our results and some concluding remarks.

\section{Diffusion and Fractional Operators}
We start by noting that the solution of the fractional diffusion equation~(\ref{PAe1b}) in the Fourier-Laplace space is
\begin{eqnarray}
\rho(k,s)=\frac{\varphi(k)}{s + s D {\cal{K}}(s) k^2}\;,
\label{solution2}
\end{eqnarray}
where $\rho(k,s)$ is the Fourier-Laplace transformation of the probability distribution $\rho(x, t)$. This result can be related to different situations depending on the choice of the kernel ${\cal{K}}(s)$. 

Within the continuous time random walk formalism and by following the works of Meztler and Klafter~\cite{15}, we can show that the waiting time $\omega(t)$ and the jump $\lambda(x)$ probability distribution associated with Eq.~(\ref{PAe1b}) are (in the Laplace and Fourier spaces) 
\begin{eqnarray}
\omega(s)=\frac{{\cal{K}}(s)/\tau_c}{1+{\cal{K}}(s)/\tau_c}
\label{waitingLaplace}
\end{eqnarray}
and
$\lambda(k)=1-k^{2} D \tau_c\,$,
where $\tau_c$ is a characteristic waiting time of the underlying continuous time random walk. We observe that the jump probability distribution is characterized by a Gaussian asymptotic behavior [$\lambda(x)\sim e^{-x^2/(2 D \tau_c)^2}$] and thus has a finite characteristic jump length, regardless of the choice for the kernel ${\cal{K}}(t)$. On the other hand, the inverse Laplace transform of the waiting time distribution is given by
\begin{eqnarray}
\omega(t)&=& \frac{1}{\tau_c}\int_{0}^{t}dt'{\cal{K}}(t') +\sum_{n=1}^{\infty}\left(-\frac{1}{\tau_c}\right)^{n}\int_{0}^{t}dt_{n}{\cal{K}}(t-t_{n})\cdots \nonumber \\ &\times& \int_{0}^{t_{4}}dt_{3}{\cal{K}}(t_{4}-t_{3})\int_{0}^{t_{2}}dt_{1}{\cal{K}}(t_{2}-t_{1}) {\cal{K}}(t_{1})\,,
\end{eqnarray}
yielding different situations that depends on ${\cal{K}}(t)$.

The choice ${\cal{K}}(t)=\delta(t)$ leads to usual diffusion and an exponential distribution for the waiting times
\begin{equation}
\omega(t)=\frac{1}{\tau_c} e^{-t/\tau_c}\,.
\label{CTRWUsual}
\end{equation}
For the fractional operator of Riemann-Liouville, we find
\begin{equation}
\omega(t)=\frac{1}{\tau_c} \left(\frac{t}{\tau}\right)^{\alpha-1}E_{\alpha,\alpha}\left[-\left(\frac{\tau}{\tau_c}\right)
 \left(\frac{t}{\tau}\right)^{\alpha}\right]\,,
\end{equation}
where $E_{\alpha,\alpha}(\dots)$ is the generalized Mittag-Leffler function~\cite{Livro1} whose asymptotic behavior is described by a power-law, $\omega(t)\sim 1/t^{1+\alpha}$ for $t\rightarrow \infty$. 
For the Atangana-Baleanu operator [Eq.~(\ref{KK33})], the waiting time distribution is given by
\begin{eqnarray}
\omega(t)=
\frac{\xi b\, \gamma }{\pi \tau_c} \sin\left(\pi\gamma\right)
 \int_{0}^{\infty}\!\!\!\!d\eta
\frac{\eta^{\gamma}e^{-\eta b (1-\alpha) t/\tau_c}}{\left(1-\eta\right)^{2}+2 \xi \left(1-\eta\right)\eta^{\gamma}\cos\left(\pi\gamma\right)+\eta^{2\gamma}}\,,
\label{KK22Inverse}
\end{eqnarray}
where $\gamma=1-\alpha$ and $\xi=(\alpha b)/(\tau_c \tau^\alpha)$. This expression is very interesting because for small times we have a stretched exponential, that is,
\begin{equation}
\omega(t)\sim E_{\alpha}\left[-\frac{\alpha}{1-\alpha}\left(\frac{t}{\tau}\right)^\alpha\right]\sim e^{-\frac{\alpha}{\Gamma(\alpha)(1-\alpha)}\left(\frac{t}{\tau}\right)^\alpha}\,,
\end{equation}
while for long times we have the same power-law behavior of the Riemann-Liouville operator. Thus, the Atangana-Baleanu operator yields a crossover between a stretched exponential and a power-law distribution.

In the case of the Caputo-Fabrizio operator, the connection with the continuous time random walk is more complex and not compatible with its standard interpretation. As we shall discuss later on, the diffusion equation associated with this operator is connected to a diffusive process with stochastic resetting~\cite{Mendez2016,Shkilev}, where the waiting time distribution is exponential.

\begin{figure}[!ht]
\centering
\includegraphics[width=0.6\textwidth]{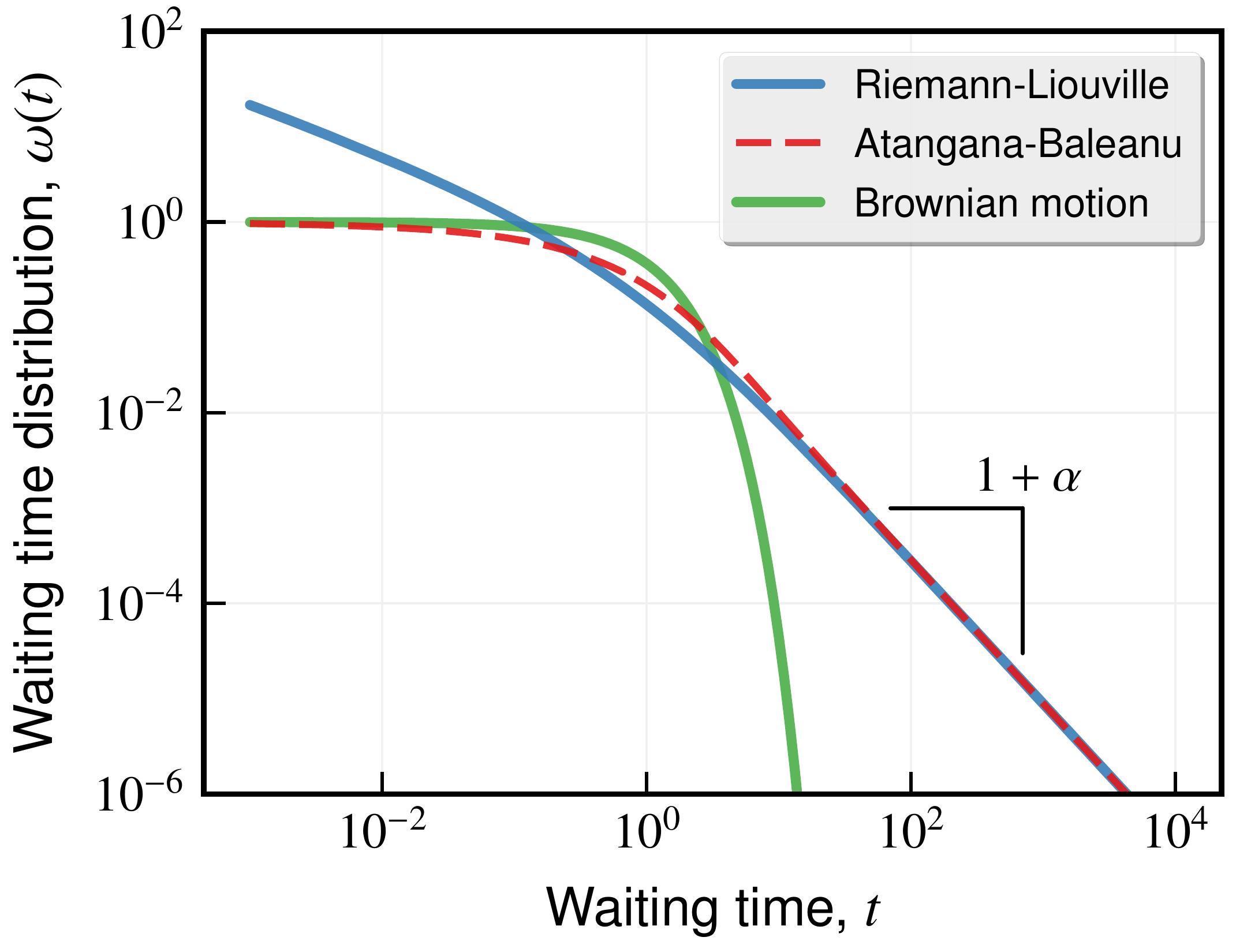}
\caption{Changes in the waiting time distribution $\omega(t)$ caused by the different forms of the kernel ${\cal K}(t)$ defining the fractional operator of Eq.~(\ref{KK}). The different curves correspond the $\omega(t)$ when choosing the kernels of Riemann-Liouville [Eq.~(\ref{KK11}), blue], Atangana-Baleanu [Eq.~(\ref{KK33}), red], and the usual (Brownian motion) case [${\cal{K}}(t)=\delta(t)$, green]. For simplicity, we have considered $\tau=\tau_c=1$ and $\alpha=1/2$. We note that the asymptotic behavior of $\omega(t)$ is a power-law for the kernels of Riemann-Liouville and Atangana-Baleanu, that is, $\omega(t)\sim 1/t^{1+\alpha}$. In the usual, we have an exponential behavior.}
\label{Figure_1}
\end{figure}

Figure~\ref{Figure_1} depicts the behavior of the waiting time distribution $\omega(t)$ for the different kernels previously-discussed. For long times, we confirm that the operators of Riemann-Liouville and Atangana-Baleanu yield the same power-law decay for $\omega(t)$. We further note that the Atangana-Baleanu operator yields a non-divergent $\omega(t)$, an interesting feature that is not observed for the singular kernel of Riemann-Liouville.

\begin{figure*}[!ht]
\centering
\includegraphics[width=1.\textwidth]{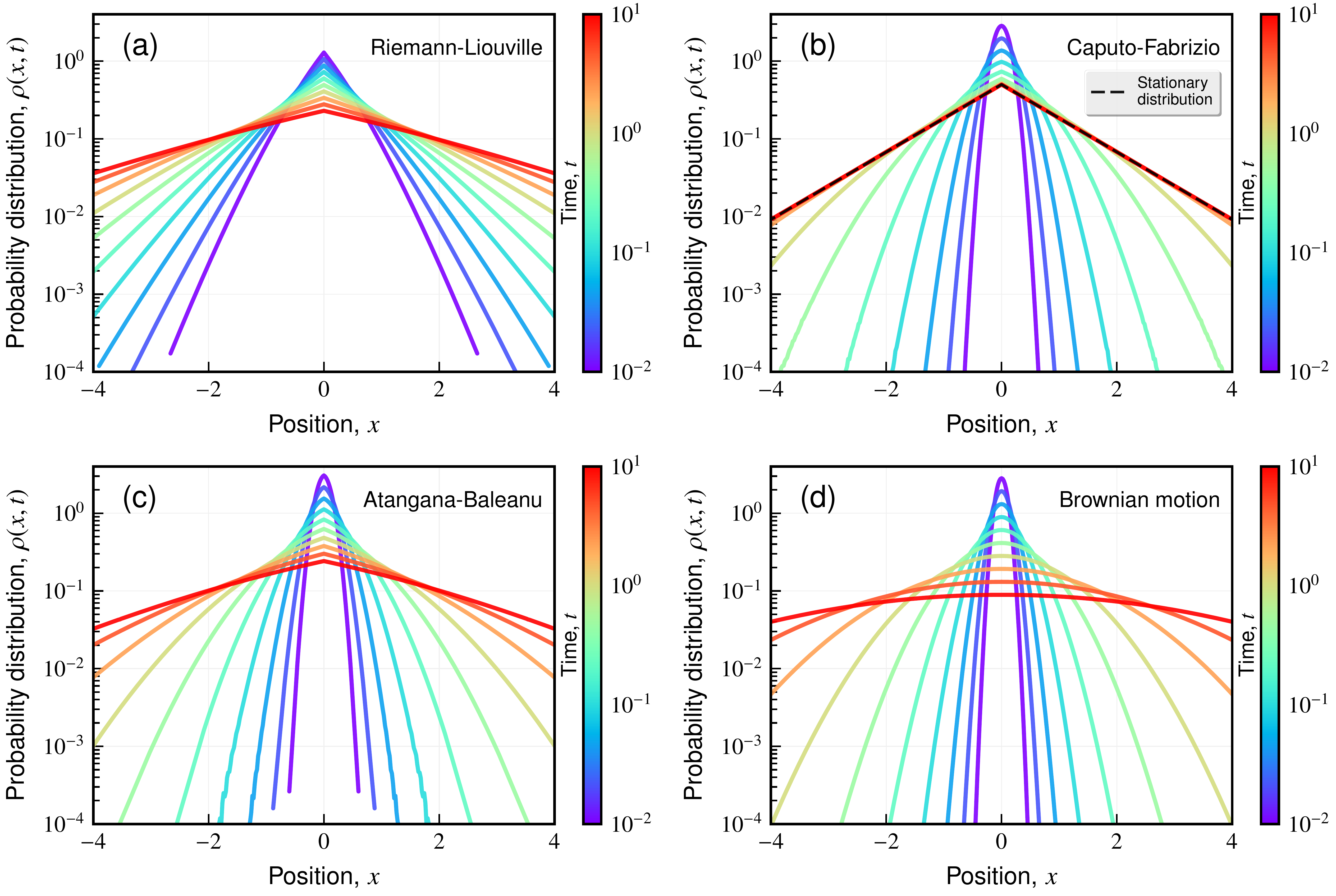}
\caption{Changes in the profile of probability distribution $\rho(x,t)$ caused by the different fractional time operators. The plots show a typical shape of $\rho(x,t)$ for different values of $t$ (indicated by the color code) when considering the operators of Riemann-Liouville (panel a), Caputo-Fabrizio (panel b), Atangana-Baleanu (panel c), and the usual case (panel d). For simplicity, we have considered $\varphi(x)=\delta(x)$, $\tau=1$, $\alpha=1/2$, and $D\,b = 1$. The dashed line in panel (b) indicates the stationary solution in the Caputo-Fabrizio [Eq.~(\ref{eq:CF_stationary})].
}
\label{FigureDistribution}
\end{figure*}

\begin{figure}[!ht]
\centering
\includegraphics[width=0.6\textwidth]{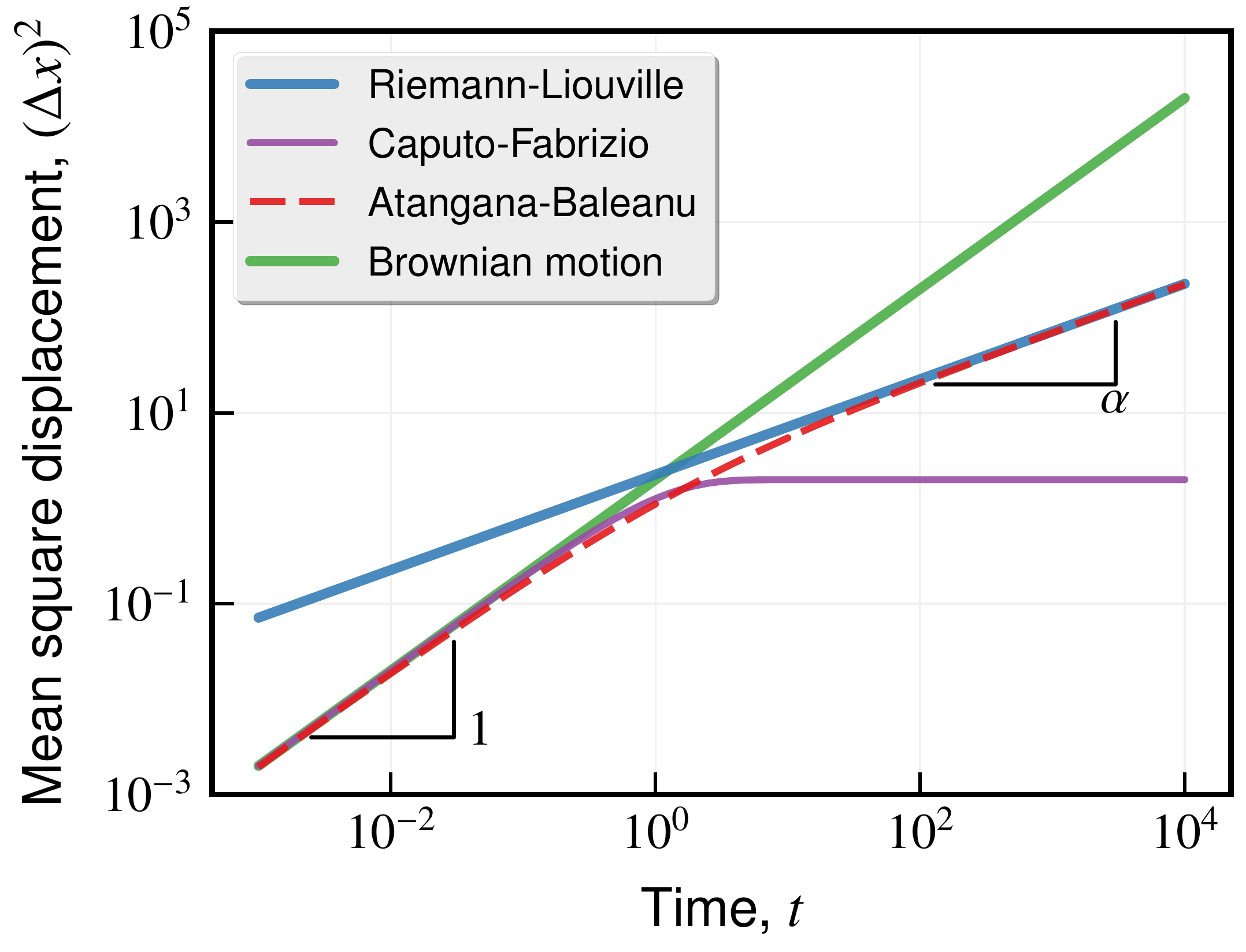}
\caption{Changes in the evolving behavior of the mean square displacement $(\Delta x)^2$ caused by the different fractional time operators. The curves show $(\Delta x)^2$ versus $t$ when considering the operators of Riemann-Liouville [blue, Eq.~(\ref{frac})], Caputo-Fabrizio [purple, Eq.~(\ref{square1})], Atangana-Baleanu [red, Eq.~(\ref{eq:MSD_AB})], and the usual case [green, $(\Delta x)^2\propto t$]. For simplicity, we have considered $\alpha=1/2$, $\tau=1$, and $D\, b=1$. It is worth noting the Atangana-Baleanu operator shows a crossover between usual (for small times) and sub-diffusion (for long time). In the Caputo-Fabrizio case, the diffusion is usual for small times and saturates for large times.}
\label{Figure_MSD}
\end{figure}

We now focus on finding the formal solutions for the fractional diffusion equation~(\ref{eq1}) when considering the three different fractional operators. These solutions are obtained by performing the inverse of Fourier and Laplace transforms of the $\rho(k,s)$ expressed in Eq.~(\ref{solution2}), where the Laplace transform of the kernel ${\cal K}(s)$ appears. In the well-known case of the Riemann-Liouville operator~\cite{15}, we have 
\begin{eqnarray}
{\cal K}(s) = \tau^{1-\alpha} s^{-\alpha}\,
\end{eqnarray}
and consequently
\begin{eqnarray}\label{eq:pdfRL}
\rho(x,t)=\int_{-\infty}^{\infty}dx'{{\cal{G}}}(x-x',t)\varphi(x')\,,
\end{eqnarray}
where the Green function is
\begin{eqnarray}\label{eq:greenRL}
{{\cal{G}}}(x,t)=\frac{1}{2|x|}{\mbox {\large{H}}}_{1,1}^{1,0} \left[ \frac{|x|}{\sqrt{D\, \tau \left(t/\tau \right)^{\alpha}}}
\left|_{\left(1,1 \right)}^{\left(1,\frac{\alpha}{2}\right)} \right. \right]\,.
\end{eqnarray}
Here ${\mbox {\large{H}}}(\dots)$ stands for the Fox H-function~\cite{Fox}. Having found the probability distribution, we can show that the mean square displacement is
\begin{eqnarray}
\left(\Delta x\right)^2= \left\langle \left(x- \left\langle x \right\rangle\right)^{2} \right\rangle =\frac{2 D\, \tau}{\Gamma(1+\alpha)} \left(\frac{t}{\tau}\right)^{\alpha}\,,
\label{frac}
\end{eqnarray}
which corresponds to the typical case of anomalous diffusion, where $\alpha<1$  represents sub-diffusion  and $\alpha\to1$ recovers the usual diffusion. The time-dependent behavior of a typical probability distribution $\rho(x,t)$ (with $\alpha=1/2$) is shown in Fig.~\ref{FigureDistribution}(a). We observe that the Riemann-Liouville operator leads to a tent-shaped distribution, whose tails are longer than the Gaussian distribution of the usual diffusion [Fig.~\ref{FigureDistribution}(d)]. Figure~\ref{Figure_MSD} shows the corresponding behavior for mean square displacement of Eq.~(\ref{frac}), which is a power-law function of the time $t$ with an exponent $\alpha$.

For the Caputo-Fabrizio operator, the Laplace transform of the kernel in Eq.~(\ref{KK22e}) is
\begin{eqnarray}
\label{KK22}
{\cal{K}}(s)&=& \frac{b}{\left(s + \frac{\alpha}{(1-\alpha) \tau}\right)}\,,
\end{eqnarray}
which substituted into Eq.~(\ref{solution2}) yields
\begin{eqnarray}
\rho(k,s)=\frac{\left(s+\frac{\alpha}{(1-\alpha) \tau}\right)\varphi(k)}{s\left(s + \frac{\alpha}{(1-\alpha) \tau} + D\, b\, k^2\right)}\;.
\end{eqnarray}
By performing the inverse Fourier and Laplace transforms, we have
\begin{eqnarray}\label{eq:pdfCF}
\rho(x,t)\!&=& \! \int_{-\infty}^{\infty}dx'{\cal{G}}(x-x',t)\varphi(x') \nonumber \\
&+&\!\frac{\alpha}{(1-\alpha) \tau}\int_{0}^{t}\!\!\!\!dt' \int_{-\infty}^{\infty}\!\!\!\!\!\!dx'{\cal{G}}(x-x',t')\varphi(x')\,,
\label{FC1}
\end{eqnarray}
where the Green function is
\begin{eqnarray}\label{eq:greenCF}
{\cal{G}}(x,t)=\frac{e^{-\frac{\alpha t}{(1-\alpha) \tau}}}{\sqrt{4\pi D\,b\, t}}\,e^{-\frac{x^{2}}{4 D\, b\, t}}\;,
\end{eqnarray}
A typical shape of this distribution is shown in Fig.~\ref{FigureDistribution}(b). We observe that this distribution is very similar to a Gaussian for small times, and exhibits a tent-shape to long times. However, differently from the distribution obtained for the Riemann-Liouville operator [Eqs.~(\ref{eq:pdfRL}) and~(\ref{eq:greenRL})], the distribution obtained from Eqs.~(\ref{eq:pdfCF}) and~(\ref{eq:greenCF}) displays a stationary behavior for $t\to\infty$, that is,
\begin{eqnarray}\label{eq:CF_stationary}
\rho(x,t\to\infty)\sim \int_{-\infty}^{\infty} dx'\varphi(x') e^{-\frac{\alpha |x-x'|}{(1-\alpha)D\,b}}\,,
\end{eqnarray}
a result that corresponds to confined diffusion. Figure~\ref{FigureDistribution}(b) also shows this stationary solution (dashed line); in particular, we observe that the shape of $\rho(x,t)$ is practically constant for $t\gtrsim 5$ in that case. This behavior also appears in the mean square displacement
\begin{eqnarray}
\left(\Delta x\right)^2= \frac{2 D\,b\, \tau (1-\alpha)}{\alpha}\left(1-e^{-\frac{\alpha t}{(1-\alpha)\tau}} \right) \;,
\label{square1}
\end{eqnarray}
which behaves linearly in time for small times and saturates in ${2 D\,b\, \tau (1-\alpha)}/{\alpha}$ for long times. Figure~(\ref{Figure_MSD}) illustrates this crossover, a common feature of systems where diffusion is confined or hindered~\cite{Manzo2015,Mo2015}. In particular, the same crossover between usual and confined diffusion is observed in simulations of diffusion with immobile obstacles or obstacles moving according to an Ornstein-Uhlenbeck process~\cite{Berry2014,Spakowitz2017}.

An intriguing feature of the diffusion equation with the Caputo-Fabrizio fractional operator is that it can be related to a diffusion with stochastic resetting~\cite{Resetting}. Indeed, we find out that the fractional diffusion equation~(\ref{PAe1b}) with the kernel of Eq.~(\ref{KK22e}) can be rewritten as 
\begin{eqnarray}
\!\!\!\frac{\partial}{\partial t}\rho(x,t)=D\,b
\frac{\partial^{2}}{\partial x^{2}}\rho(x,t)-  \frac{D\,b\,\alpha}{(1-\alpha)\tau}\int_{0}^{t}dt'
e^{-\frac{\alpha(t-t')}{(1-\alpha)\tau}}\frac{\partial^{2}}{\partial x^{2}}\rho(x,t')\;.
\label{KP1}
\end{eqnarray}
Also, by integrating both sides of the fractional diffusion equation~(\ref{PAe1b}), we obtain
\begin{eqnarray}
D\, b\int_{0}^{t}dt' e^{-\frac{\alpha(t-t')}{(1-\alpha)\tau}}\frac{\partial^{2}}{\partial x^{2}}\rho(x,t')=\rho(x,t)-\varphi(x)\,,
\end{eqnarray}
which after substituting into Eq.~(\ref{KP1}) yields
\begin{eqnarray}
\frac{\partial}{\partial t}\rho(x,t) = D\,b
\frac{\partial^{2}}{\partial x^{2}}\rho(x,t) 
- \frac{\alpha}{(1-\alpha)\tau}\left[\rho(x,t)-\varphi(x)\right]\,.
\label{KP}
\end{eqnarray}
Equation~(\ref{KP}) with $\varphi(x)=\delta(x-x_0)$ is the same obtained by Evans and Majumdar~\cite{Resetting} when studying a random walker whose position is redefined to the position $x_0$ with a rate $r=\frac{\alpha}{(1-\alpha)\tau}$. Thus, the fractional exponent $\alpha$ (as well as the parameter $\tau$) in the fractional diffusion equation of Caputo-Fabrizio can be related to a well-defined physical quantity (resetting rate). 

Also, the mean square displacement of Eq.~(\ref{square1}) is analogous to results obtained from a random walk description of a diffusive process with stochastic resetting, subjected to an exponential waiting time distribution~\cite{Mendez2016,Shkilev}. As discussed in these works, a suitable continuous time random walk formulation is established by considering a density of particles ${\cal{J}}(x,t)$ whose dynamics is governed by   
\begin{eqnarray}
{\cal{J}}(x,t)&=&\delta(t)\delta(x)+
{r}\delta(x)\int_{0}^{t}\!\!dt'\omega(t'){\cal{J}}(x,t-t') \nonumber \\ 
&+&(1-{r})\int_{0}^{t}\!\!dt'\psi(x',t'){\cal{J}}(x-x',t-t')\,,
\label{CTRWModificado}
\end{eqnarray}
when particles start the random walk at the origin ($x=0$) with $\rho(x,t)=\int_{0}^{t}\Phi(t'){\cal{J}}(x,t-t')dt'$ and $\Phi(t)=\int_{t}^{\infty}\omega(t')dt'$. In Eq.~(\ref{CTRWModificado}), $r$ is a resetting rate, $\psi(x,t)$ is joint distribution of jump length and waiting time, $\lambda(x)=\int_{0}^{\infty}dt\psi(x,t)$ is the jump length distribution, and $\omega(t)=\int_{-\infty}^{\infty}dx\psi(x,t)$ is waiting time distribution. By considering $\lambda(x)$ Gaussian and $\omega(t)$ exponentially distributed, we can show that this formalism leads to Eq.~(\ref{KP}). It is worth remarking that by comparison with this framework, we can infer that the diffusion equation with the Caputo-Fabrizio operator leads to the same waiting time distribution of the usual diffusion, that is, an exponential.

Finally, for the Atangana-Baleanu operator, the Laplace transform of the kernel in Eq.~(\ref{KK33}) is
\begin{eqnarray}
\label{KK22L}
{\cal{K}}(s)= \frac{b s^{\alpha-1}}{\left(s^{\alpha} + \frac{\alpha}{(1-\alpha)\tau^\alpha}\right)}\,,
\end{eqnarray}
which substituted into Eq.~(\ref{solution2}) yields
\begin{eqnarray}
\rho(k,s)=\frac{\left(s^{\alpha}+\frac{\alpha}{(1-\alpha)\tau^\alpha}\right)\varphi(k)}{s^{\alpha}\left(s+\frac{\alpha s^{1-\alpha}}{(1-\alpha)\tau^\alpha} + D\,b\, k^2\right)}\,,
\end{eqnarray}
the solution for the fractional diffusion equation~(\ref{PAe1b}) in the Fourier-Laplace space. By evaluating the inverse Fourier and Laplace transforms, we obtain
\begin{eqnarray}
\rho(x,t) &=& \int_{-\infty}^{\infty}dx'{{\cal{G}}}(x-x',t)\varphi(x')\nonumber \\ &+& \frac{\alpha}{\Gamma(\alpha)(1-\alpha)\tau^\alpha}\int_{0}^{t}\frac{dt'}{(t-t')^{1-\alpha}}\int_{-\infty}^{\infty}dx'\varphi(x'){{\cal{G}}}(x-x',t'),
\end{eqnarray}
where the Green function is
\begin{eqnarray}
{{\cal{G}}}(x,t)= \frac{e^{-\frac{x^{2}}{4 D\,b\, t}}}{\sqrt{4\pi D\,b\, t}}+ \frac{1}{|x|}\sum_{n=1}^{\infty}\frac{\left(-\frac{\alpha}{1-\alpha} \right)^n}{\Gamma(1+n)} \left(\frac{t}{\tau}\right)^{n\alpha}
{\mbox {\large{H}}}_{2,2}^{2,0} \left[ \frac{x^{2}}{\sqrt{D\,b\,t}}
\left|_{\left(1,2 \right),\left(1+n,1 \right)}^{\left(1+\alpha n,1\right),(1,1)} \right. \right].
\label{green2}
\end{eqnarray}
Once again, ${\mbox {\large{H}}}(\dots)$ stands for the Fox H-function~\cite{Fox}. We can also show that for $|x|\rightarrow \infty$, Eq.~(\ref{green2}) is approximated by
\begin{eqnarray}
\label{Asymp}
{{\cal{G}}}(x,t) \approx  \frac{1}{\sqrt{4\pi D\,b\, t}}e^{-f(x,t)}\,,
\end{eqnarray}
where
\begin{eqnarray}\label{Asymp2}
f(x,t) = \frac{x^{2}}{4 D\,b\, t}+ \frac{\alpha}{1-\alpha} \left(\frac{t}{\tau}\right)^{\alpha}\left(\frac{x^{2}}{4 D\, b\, t}\right)^{1-\alpha}\,.
\end{eqnarray}
A typical behavior for the distribution $\rho(x,t)$ for this operator is shown in Fig.~\ref{FigureDistribution}(c). Similarly to the Caputo-Fabrizio operator, the profile of $\rho(x,t)$ resembles a Gaussian for small times, while exhibits a tent-shape for long times. However, the distribution does not have a stationary solution for the Atangana-Baleanu operator. This crossover between two behaviors for $\rho(x,t)$ is also present in Eq.~(\ref{Asymp2}), and can be better quantified by analyzing the mean square displacement. For this operator, we have
\begin{eqnarray}\label{eq:MSD_AB}
\left(\Delta x\right)^2 = 2 D\, b\, t E_{\alpha,2}\left[-
\frac{\alpha}{1-\alpha}\left(\frac{t}{\tau}\right)^\alpha\right]\;,
\end{eqnarray}
where $E_{\alpha,2}(\dots)$ is the generalized Mittag-Leffler function~\cite{Livro1}. By considering the asymptotic limits of this function, we can show that $\left(\Delta x\right)^2\sim t$ for small times, and $\left(\Delta x\right)^2\sim t^{1-\alpha}$ for long times. 

This crossover between usual and sub-diffusion is present in several biological systems~\cite{Skalski2000,Sawadac2001,Mieruszynsky2015,ribeiro2014investigating,Ribeiro2016} and is also illustrated in Fig.~\ref{Figure_MSD} for $\alpha=1/2$. A similar situation appears in simulations of diffusion with obstacles moving according to a usual random walk~\cite{Berry2014,Spakowitz2017}, where the same crossover between usual and sub-diffusion with $\alpha=1/2$ is observed. It is worth mentioning that crossovers between diffusive regimes can also be described by generalized Langevin equations~\cite{Tateishi2012} and fractional (with the Riemann-Liouville operator) Kramers equations~\cite{Barkai1}, among other approaches~\cite{DifferentRegimes2,DifferentRegimes1}. In particular, the usual Langevin equation~\cite{Bian2016} predicts a crossover between ballistic and usual diffusion, which has been experimentally observed only in 2011~\cite{HuangRaizen2011}. However, the diffusion equation in terms of these new operators lead to these crossovers without explicitly considering external forces, inertial effects, and reaction terms.

The fractional diffusion equation with the Atangana-Baleanu operator can be further related to fractional derivatives of distributed order as proposed by Caputo~\cite{caputo1995mean,caputo2001distributed} and worked out in Refs.~\cite{DifferentRegimes2,DifferentRegimes1}, that is,
\begin{equation}\label{eq:frac_ord_diff}
\int_0^1\!\! d\nu\, w(\nu)\frac{\partial^{\nu}}{\partial t^{\nu}} \rho(x, t) = D \frac{\partial^2}{\partial x^2} \rho(x,t)\,,
\end{equation}
where $w(\nu)$ is the distribution of the fractional order exponent $\nu$ and 
\begin{equation}
\frac{\partial^\nu}{\partial t^\nu} \rho(x,t) = \frac{1}{\Gamma\left(1-\nu\right)}\int_{0}^{t}\frac{dt'}{(t-t')^{\nu}}\frac{\partial}{\partial t'}\rho(x,t')
\end{equation}
is the fractional time derivative of Caputo. Indeed, by substituting the kernel of Eq.~(\ref{KK33}) into Eq.~(\ref{PAe1b}) and taking the Laplace transform, we have
\begin{eqnarray}
\!\!\!\!\!\!\!\!\!s\rho(x,s)-\rho(x,0)= D\,b \frac{s^{1-\alpha}}{s^{1-\alpha}
+ \frac{\alpha}{(1-\alpha)\tau^\alpha}}\frac{\partial^{2}}{\partial x^{2}}\rho(x,s)\,,
\end{eqnarray}
which can be rewritten as
\begin{eqnarray}
s\rho(x,s) -\rho(x,0) \frac{\alpha}{(1-\alpha)\tau^\alpha}s^{-\alpha}\left[s\rho(x,s)-\rho(x,0)\right]
= D\,b\,\frac{\partial^{2}}{\partial x^{2}}\rho(x,s)\,.
\end{eqnarray}
By calculating the inverse Laplace transform of the previous equation, we find
\begin{eqnarray}
\frac{\partial}{\partial t}\rho(x,t) + \frac{\alpha}{\Gamma(\alpha)(1-\alpha)\tau^\alpha}\int_{0}^{t}\frac{dt'}{(t-t')^{1-\alpha}}\frac{\partial}{\partial t'}\rho(x,t') = D\,b\,\frac{\partial^{2}}{\partial x^{2}}\rho(x,t) \;,
\label{KP2}
\end{eqnarray}
which can also be written as
\begin{eqnarray}\label{eq:difcaputo}
\frac{\partial}{\partial t}\rho(x,t) + \frac{\alpha}{(1-\alpha)\tau^\alpha} \frac{\partial^{1-\alpha}}{\partial t^{1-\alpha}}\rho(x,t) 
= D\,b\,\frac{\partial^{2}}{\partial x^{2}}\rho(x,t)\,.
\end{eqnarray}
We note that Eq.~(\ref{eq:difcaputo}) is a special case of Eq.~(\ref{eq:frac_ord_diff}) with $w(\nu) = \delta(\nu -1) + \frac{\alpha}{(1-\alpha)\tau^\alpha} \delta(\nu +\alpha -1)$. Analogously to results reported here, the solutions of Eq.~(\ref{eq:difcaputo}) are also characterized by two diffusive regimes~\cite{DifferentRegimes2,DifferentRegimes1}.

\section{Discussion and Conclusions}

We presented a detailed investigation of the changes in the fractional diffusion equation when the well-established Riemann-Liouville operator is replaced by the recently-proposed operators of Caputo-Fabrizio and Atangana-Baleanu. These changes are summarized in Table~\ref{tab:summary}. Within the context of the continuous time random walk, we verified that these new fractional operators modify the behavior of the waiting time distribution. In the Caputo-Fabrizio case, we found that the waiting time distribution is described by an exponential distribution; while the Atangana-Baleanu operator yields a distribution that decays as a stretched exponential for small times and as a power-law (with the same exponent of the Riemann-Liouville operator) for long times. 

\begin{table}[!ht]
\caption{Summary of the changes caused by the different fractional operators on the diffusion equation~(\ref{eq1}).}
\label{tab:summary}
\centering
\small
\tabcolsep=0.06cm
\begin{tabular}{l|l|l|l}
\hline
\thead{\bf Fractional\\\bf Operator} & \thead{\bf Waiting time \\\bf distribution} & \thead{\bf Mean square\\\bf displacement} & \thead{\bf Probability\\\bf distribution} \\
\hline
\thead{Riemann-\\Liouville} & \thead{power-law} & \thead{power-law and\\scale-invariant} & \thead{non-Gaussian} \\
\hline
\thead{Caputo-\\Fabrizio} & \thead{exponential} & \thead{crossover from\\usual to confined\\diffusion} & \thead{crossover from\\Gaussian to\\ non-Gaussian\\with steady state} \\
\hline
\thead{Atangana-\\Baleanu} & \thead{crossover from\\stretched exp.\\to power-law} & \thead{crossover from\\usual to\\ sub-diffusion} & \thead{crossover from\\Gaussian to\\ non-Gaussian} \\
\hline
\end{tabular}
\end{table}

We obtained the exact solutions of the fractional diffusion equation and the time dependence of the mean square displacement when considering these different fractional operators. Our results reveal that these new operators lead to non-Gaussian distributions and different diffusive regimes depending on the time scale. For the Caputo-Fabrizio operator, the probability distribution $\rho(x,t)$ displays a stationary state as well as saturated diffusion for long times. This is a remarkable feature because the fractional diffusion equation is solved without external forces and subjected to the free diffusion boundary conditions. For the Atangana-Baleanu operator, we found a crossover between two diffusive regimes: a usual for small times and a sub-diffusive for long times, a feature observed in several empirical systems.

By properly manipulating the fractional diffusion equations, we demonstrated that the results obtained with these new fractional operators could be connected with other diffusive models. The fractional diffusion equation with the Caputo-Fabrizio operator recovers a diffusive process with stochastic resetting, where the fractional order exponent is directly related to the resetting rate. Also, the equation with the Atangana-Baleanu operator can be associated with a fractional diffusion equation with derivatives of distributed order. Our results thus show that these new fractional operators are a simple and efficient way for incorporating different structural aspects into the system, opening new possibilities for modeling and investigating the interplay of different physical mechanisms of anomalous diffusion.

\section*{Data Availability}
\noindent There no datasets involved in this work. All graphs were produced by using matplotlib, an open source plotting library for the Python programming language.

\section*{Competing Interests}
\noindent We have no competing interests.

\section*{Authors' Contributions}
\noindent All the authors were involved in the design of the study, calculations, preparation of the manuscript, and gave final approval for publication.

\section*{Funding}
\noindent AAT thanks the financial support of the CNPq under Grant No. 462067/2014-9. HVR thanks the financial support of the CNPq under Grant No. 440650/2014-3 and CAPES. EKL thanks the financial support of the CNPq under Grant No. 303642/2014-9.

\end{document}